\documentclass[aps,prb,amssymb,twocolumn,showpacs]{revtex4}
\usepackage{epsfig}
\begin{document}

\title{Domain Wall Renormalization Group Study of
XY Model with Quenched Random Phase Shifts}
\author{N. Akino$^{1}$ and J.M. Kosterlitz$^{2}$}
\affiliation{$^{1}$Institut f\"ur Physik, WA 331, Johannes Gutenberg-Universit\"at,
         D-55099 Mainz, Germany}
\affiliation{$^{2}$Department of Physics, Brown University,
         Box 1843, Providence, RI 02912, USA}
\date{\today}
\begin{abstract}
The $XY$ model with quenched random disorder is studied by a zero
temperature domain wall renormalization group method in $2D$ and $3D$.
Instead of the usual phase representation we use the charge (vortex)
representation to compute the domain wall, or defect, energy.
For the gauge glass corresponding to the maximum disorder
we reconfirm earlier predictions that
there is no ordered phase in $2D$ but an ordered phase can exist in $3D$
at low temperature. However, our simulations yield spin stiffness exponents
$\theta_{s} \approx -0.36$ in $2D$ and $\theta_{s} \approx +0.31$ in $3D$,
which are considerably larger than previous estimates
and strongly suggest that the lower critical dimension is less than
three. For the $\pm J$ $XY$ spin glass in $3D$, we obtain a spin stiffness
exponent $\theta_{s} \approx +0.10$ which supports the existence of
spin glass order at finite temperature in contrast with previous
estimates which obtain $\theta_{s}< 0$.
Our method also allows us to study renormalization group flows of both the
coupling constant and the disorder strength with length scale $L$.
Our results are consistent with recent analytic and numerical studies
suggesting the absence of a re-entrant transition in $2D$ at low temperature.
Some possible consequences and connections with real vortex systems are discussed.
\end{abstract}
\maketitle

\section{\bf Introduction}
\label{sec:intro}

  The $XY$ model with quenched random phase shifts
as a model for a superconducting glass has been intensively
investigated over the last decade, focusing on the
so-called gauge glass model which corresponds to the case
with maximal disorder.
Since a transport current exerts a force on a flux lattice, it tends to
move in response which causes dissipation of the current.
The existence of disorder, which destroys the flux lattice structure, is
essential to pin the vortices in order for a superconducting phase
to exist in a high $T_{c}$ superconductor \cite{vg-fisher,vg-blatter,vg-natter}.
Although there exists no proof whether or not the gauge glass
and vortex glass are in the same universality class,
it is of interest as the simplest model of a disordered superconductor
and is still not understood despite all the effort expended on it.
\newline\indent
  From numerical\cite{gg-hs,gg-rtyf,gg-g,gg-fty,gg-ks,gg-mg}
and experimental \cite{experiment1}
studies, it is believed that the gauge glass has no ordered phase at any
finite
temperature in two dimensions.
In three dimensions, numerical domain wall renormalization group (DWRG)
studies\cite{DWRG1,DWRG2} indicate
that the lower critical dimension seems to be close to three.
However the situation is less conclusive, since the simulations are
limited to small system sizes. Finite temperature Monte Carlo studies
yield a transition temperature $T_{c}/J \sim
O(1)$\cite{gg-hs,gg-rtyf,gg-wy} which is difficult to
reconcile with DWRG studies\cite{gg-rtyf,gg-g,gg-ks,gg-mg}
as these studies imply
that the lower critical dimension for superconducting glass order
is close to three. Experimentally there is also some
evidence for a finite temperature phase transition to
a superconducting glass phase \cite{experiment2,experiment3}.
\newline\indent
  The Hamiltonian of the XY model with random quenched disorder can be
written as
\begin{equation}
\label{eq:h}
H=\sum_{<ij>}V(\theta_{i}-\theta_{j}-A_{ij})
\end{equation}
where $V(\phi)$ is an even, $2\pi$ periodic function of $\phi$ with a
maximum
at $\phi=\pi$ and minimum at $\phi=0$, usually taken as
$V(\phi_{ij})=-J_{ij}{\rm cos}(\phi_{ij})$.
The sum is over all nearest neighbor pairs of sites and the coupling
constants, $J_{ij}$, are assumed uniform, $J_{ij}=J>0$.
The random bond variables $A_{ij}$, which are responsible
for the randomness and frustration, are taken to be independent and uniformly
distributed in $(-\alpha\pi,\alpha\pi]$ with $0 \le \alpha \le 1$.
For a gauge glass, $\theta_{i}$ is the phase of the superconducting
order parameter at site $i$ of a square lattice in $2D$ and
a simple cubic lattice in $3D$.
The random bond variables $A_{ij}$ are taken to correspond to maximal
disorder with $\alpha = 1$. An external field applied to an
extreme type II superconductor induces a uniform
component $A^{0}_{ij}=(2\pi/\Phi_{0})\int_{i}^{j}{\bf A}\cdot d{\bf l}$
where ${\bf A}$ is the vector potential of the applied field
and $\Phi_{0}=hc/2e$ is the quantum of flux. In this work, we take
$A^{0}_{ij}=0$, corresponding to zero applied field. Unless explicitly
stated, we consider an unscreened system with $\alpha = 1$
corresponding to maximal disorder.
The Hamiltonian of Eq. (\ref{eq:h}) also describes
the $XY$ magnet with random Dzyaloshinski-Moriya interactions
\cite{xy-magnet}
and also a Josephson junction array with positional disorder. \cite{gk1,gk2}
These studies \cite{xy-magnet,gk1,gk2} showed that the existence of
weak disorder $(\alpha \ll 1)$ does not destroy an ordered phase at
intermediate
temperature but predict a re-entrant transition to a disordered phase at
low temperature in two dimensions.
However, recent analytic \cite{xyrdm-nsk,xyrdm-kn,xyrdm-s,xyrdm-cf}
and numerical \cite{gg-ks} studies suggest the absence of
a re-entrant transition and that
there exists an ordered phase for $T<T_{c}(\alpha)$
when $\alpha < \alpha_{c}$.
\newline\indent
When the random bond variables
$A_{ij}$ are restricted to $0$ or $\pi$ with equal probabilities,
this model reduces to the $\pm J$ $XY$ spin glass,
which is believed to be in a different universality class
due to the additional reflection symmetry \cite{villain-vsg}
which is absent in the case of uniformly or Gaussian distributed
$A_{ij}$. An $XY$ spin glass may have
both spin and chiral glass order associated with rotational and
reflection symmetry, respectively. It has been suggested that, in $2D$
and $3D$, spin and chiral variables decouple at long distances and order
independently \cite{sg-mc-kt,sg-dw-kt,sg-dw-k,sg-dw-mg},
and the lower critical dimensions are $d_{l} \ge 4$ for spin glass
order and $d_{l}<3$ for chiral glass order. However, the decoupling
scenario contradicts the analytic studies on a ladder lattice
\cite{nhm}, on a tube lattice \cite{tnh}, and on a $2D$ lattice with
a special choice of disorder \cite{nh}. Recent numerical simulations
\cite{sg-dw-mg} also suggests $d_{l}$ for a spin glass order may be
close to three.
\newline\indent
In this paper, we re-investigate the possibility of an ordered phase at
small but finite temperature $T$ by a numerical
domain wall renormalization group (DWRG)\cite{DWRG1,DWRG2}, or defect
energy scaling.
The domain wall or defect energy of the system is computed by using
the Hamiltonian in the Coulomb gas (vortex) representation,
which is more convenient for numerical work
as it eliminates spin wave contributions to the energy.
Although the conventional DWRG method can handle
only the scaling of the coupling constant $J(L)$ at scale $L$,  which is
proportional to the domain wall or defect energy, our method enables us
to study the flows of both the coupling constant and the disorder
strength, $A(L)$, at length scale $L$ \cite{gg-ks}.
We apply this to the case of general disorder strength, $0 \le \alpha \le 1$.
The outline of the paper is as follows.
In Section \ref{sec:strategy} we discuss the DWRG method and also our strategy.
In Section \ref{sec:cg}, we explicitly perform
the transformation of the $3D$ Hamiltonian of Eq. (\ref{eq:h})
from the phase to the Coulomb gas representation.
Our numerical method is explained in Section \ref{sec:method}.
Finally we discuss our numerical results in Section \ref{sec:results}
and give a brief discussion of some of the effects of weak disorder,
$\alpha< 1$, and of finite screening of vortex - vortex interactions.

\section{\bf Strategy}
\label{sec:strategy}

The general idea behind a DWRG is to compute, analytically or numerically,
the energy $\Delta E(L)$ of a domain wall in a system of linear size $L$
and fit this to a finite size scaling form
\begin{equation}
\Delta E(L)\sim L^{\theta}
\label{eq:fss1}
\end{equation}
where $\theta$ is a stiffness exponent, whose sign is of fundamental
importance. If $\theta <0$, $\Delta E(L)$ vanishes in the thermodynamic
limit. The energy of the domain wall or defect excitation vanishes which
implies that, for $T>0$, the probability of the defect
$P_{L}\sim e^{-\Delta E(L)/kT}\rightarrow 1$ as $L\rightarrow\infty$. This
in turn implies that the density of such defects is finite when $T>0$ and
there will be no resistance to an infinitesimal applied force and the system
has no order. This is analogous to the vanishing of the shear modulus in
a liquid, the superfluid density in a superfluid or superconductor
and the spin stiffness constant in an isotropic magnet when $T>T_{c}$. On the
other hand, if $\theta>0$, such defects will have zero probability when
$L=\infty$ and the system will have finite stiffness and will be ordered
at sufficiently small $T>0$.
\newline\indent
In a uniform system without disorder, the definition of the energy of a domain
wall of size $L$, $\Delta E(L)$, is intuitively obvious. The first step is to find
the ground state (GS) energy of a system of size $L$, which requires applying boundary
conditions (BC) which are compatible with the GS configuration. For a
ferromagnet, this is straightforward to implement as the GS configuration
is known to be one with all spins parallel and periodic BC are compatible
with this. To impose a spin domain wall perpendicular to the ${\bf\hat x}$
direction, one simply changes the BC to antiperiodic along ${\bf\hat x}$ and
periodic in the other $d-1$ directions. Then it immediately follows that
\begin{equation}
\Delta E(L)= E_{ap}(L)-E_{p}(L)\sim L^{d-n}
\end{equation}
where $n=1$ for an Ising model and $n=2$ for a system with a continuous
symmetry such as $XY$ and Heisenberg models.
\newline\indent
One would like to use the same strategy for {\it random} systems described by
Eq. (\ref{eq:h}), as suggested by Anderson for Ising spin glasses\cite{pwa}.
However, it is not so clear how to proceed because, for a particular sample
(realization of disorder), neither the GS configuration nor compatible BC
is known so computing the defect energy $\Delta E(L)$ is
problematical. Assuming $\Delta E(L)$ can be calculated, the stiffness
exponent is defined by the scaling {\it ansatz}
\begin{equation}
\langle \Delta E(L) \rangle \sim L^{\theta}
\label{eq:fss}
\end{equation}
where $\langle \cdots \rangle$
denotes an average over realizations of disorder. To our knowledge, it is
not known how to calculate {\it analytically} either the GS energy
$E_{0}(L)$ or the energy $E_{D}(L)$ of the system containing a defect
{\it relative to this GS} which means that one must proceed numerically.
A number of conceptual and technical difficulties are apparent. The first,
and most important, is the technical problem of computing the energy
difference $\Delta E(L)$ between the energies of the system subject to two
different BC. We ultimately want the disorder averaged defect energy
$\langle \Delta E(L) \rangle = \langle E_{a}(L) - E_{b}(L) \rangle$
where $E_{a}(L)$ is the lowest energy of a particular sample subject to BC
denoted by $a$ and $E_{b}(L)$ with BC denoted by $b$. We need the individual
energies $E_{a}(L)$ and $E_{b}(L)$ essentially {\it exactly} because the
uncertainty in $\langle \Delta E(L)\rangle $ must be kept as small as possible.
Also, to our knowledge, there is no proof that the scaling ansatz of Eq. (\ref{eq:fss})
is a correct description and, even if it is, the only thing we can be sure of is
$\theta\leq (d-2)/2$. All results are based on fitting data to the scaling form of
Eq. (\ref{eq:fss}) so one is attempting {\it both} to verify the scaling ansatz and to
estimate a numerical value of $\theta$. For any conclusion to be believable, the data
must have both very small errors and fit Eq. (\ref{eq:fss}) extremely well. The first
requirement of highly accurate data is the most important as the estimate of
$\theta$ depends on this. Assuming that $E_{a}(L)$ and $E_{b}(L)$ can be
determined exactly for each sample, then $\Delta E(L)$ is also known exactly for each
sample and the errors in $\langle \Delta E(L) \rangle$ are $O(N^{-1/2}L^{d-1})$ where
$N$ is the number of samples of size $L$ in $d$ dimensions. If the energy minima
$E_{a,b}(L)$ are not found exactly, a crude estimate of the errors in
$\Delta E(L)$ is $O(N^{-1/2}L^{d})$ but this is certainly too low as failure of the
algorithm to find the true minima because of being trapped in a metastable state of
energy $E > E_{0}$ will cause systematic errors of unknown magnitude.
Empirically, we find that this can readily cause errors larger than
$\langle \Delta E(L) \rangle$ which makes the data point useless. This is most
likely to happen for large $L$ because the CPU time required grows
uncontrollably, as do the errors, so the large $L$ data becomes unreliable.
This technical difficulty limits the accessible sizes $L$ to small values
as one must keep errors in individual data points small.
\newline\indent
We are forced to conclude that the accessible sizes $L$ are limited by the
necessity of finding essentially exact global energy minima of each of a
number $N$ of samples subject to certain, yet to be defined, BC. To our
knowledge, there is no algorithm applicable to the systems of interest
which will find exact minima in polynomial time, such as the branch and
cut algorithm\cite{rieger} for the $2D$ Ising spin glass or numerically exact
combinatorial optimization algorithms \cite{combin} for gauge and vortex glass
models in the infinite screening limit, so we have to live with the fact that
our problem is NP complete and the required CPU time explodes as $L$ increases.
We use simulated annealing\cite{anneal1,anneal2} to estimate the lowest energies,
which seems considerably more efficient than simple quenching to $T=0$, but we are
unable to go beyond $L=7$ in $3D$ and $L=10$ in $2D$. We wish to
extract the stiffness exponent $\theta$ from the scaling ansatz of
Eq. (\ref{eq:fss}) with a single power law and this makes sense only if the
errors on individual data points are very small and the fit to the assumed
scaling form is extremely good. In our opinion, the only sensible strategy
is to obtain very accurate estimates of $\langle \Delta E(L) \rangle$
for the limited sizes $L$ which are feasible for the computer power available.
\newline\indent
In the phase representation of Eq. (\ref{eq:h}), the configuration space to
be searched for the global energy minima $E_{a,b}(L)$ is rather large as
the phases $\theta_{i}\in (0,2\pi]$ are continuous variables. Searching
this space in finite time is not feasible as most of the
allowed configurations of the $\theta_{i}$ are not even local energy minima.
It is well known that random $XY$ models with a Hamiltonian of Eq. (\ref{eq:h}) can be
written in a Coulomb gas (CG) or vortex representation via a duality
transformation\cite{duality1,duality2,nh} which leaves the partition
function invariant. This expresses the Hamiltonian in terms of charge or
vortex configurations which are {\it{already}} local energy minima
\cite{kt,jmk}. Thus, a
reformulation of the Hamiltonian of Eq. (\ref{eq:h}) as a CG performs a
partial minimization. A further minimization of the CG Hamiltonian corresponds to
searching the much smaller space of local minima. Reformulating the problem of
Eq. (\ref{eq:h}) including the BC in CG language is clearly a worthwhile exercise
as it dramatically reduces the number of configurations over which we have to minimize,
despite introducing long-ranged Coulomb interactions between vortices. The
transformation is carried out in Section \ref{sec:cg} for the model of Eq. (\ref{eq:h}) in $3D$.
\newline\indent
The final problem is to define what is meant by a domain wall and the BC
needed to induce a wall in a finite system of size $L$ in $d$ dimensions. We
imagine the system of Eq. (\ref{eq:h}) on a torus in $2D$ or a hypertorus in
$3D$, which corresponds to imposing periodic BC in the phases
$\theta_{{\bf i}+L{\bf\hat e}_{\mu}} = \theta_{\bf i}$ where ${\bf\hat e}_{\mu}$
is a unit vector in the direction $\mu=x,y,\cdots,d$ and
${\bf i} = (i_{x},\cdots,i_{d})$ with $i_{\mu}=(1,2,\cdots,L)$. The phases
at corresponding sites $(i,j)$ on opposite faces are coupled by some interaction
$\tilde{V}(\theta_{i},\theta_{j},A_{ij})$ which may be regarded as defining the BC.
In principle, the GS is obtained by minimizing the energy with respect to
the $L^{d}$ bulk variables $\theta_{i}$ and all forms of $\tilde{V}$. This
program is beyond our ability and we restrict ourselves to those $\tilde{V}$
which induce a spin or chiral defect, which are related to the continuous
and reflection symmetry respectively. To impose a spin defect, we choose
$\tilde{V} = V(\theta_{i}-\theta_{j}-A_{ij})$.
The plaquettes between the opposite faces are indistinguishable
from the others and play no special role. We therefore keep fixed the
frustrations $f_{\bf r} = \sum_{P({\bf r})}A_{ij}/2\pi$ where the
sum is over the bonds in a clockwise direction of the elementary plaquette
centered at ${\bf r}$. We still have the freedom to add $\Delta_{\mu}$ to every
bond in the $\mu$ direction between opposite faces which imposes a
global twist $\Delta_{\mu}$ in the phase round a loop circling the hypertorus
in the direction $\mu$. This is equivalent to a gauge transformation
$A_{ij}\rightarrow A_{ij}+\Delta_{\mu}/L$ on every bond $ij$ in the
direction $\mu$. The lowest energy, $E_{0}(\Delta_{\mu})$, is $2\pi$ periodic
in $\Delta_{\mu}$ with a minimum at some $\Delta^{0}_{\mu}$ which depends on
the sample. To induce a spin domain wall normal to ${\bf x}$, one changes the
twists from their best twist (BT) values
$\Delta^{0}_{\mu}\rightarrow \Delta^{0}_{\mu} + \pi\delta_{\mu ,x}$. The
minimum energy subject to this constraint gives $E_{sD}(L)$, the energy of the
system of size $L$ containing an extra spin defect. Note that $E_{sD}\geq E_{0}$
for every sample but $E_{0}$ is not necessarily the absolute minimum as some
other functional form of $\tilde{V}$ may give a lower energy. However, even if
$E_{0}$ is not the true GS energy but is the energy of a state with some
excitation from the GS, this method of inducing a spin defect ensures that any
excitation in the BT configuration will also be present in the state with an
extra spin domain wall so that
$\Delta E_{s}^{BT}(L)\equiv E_{sD}(L) - E_{0}(L) \geq 0$ is not affected by
these. It is convenient, but not necessary, to define the spin defect
energy by a twist of $\pi$ from the BT value $\Delta_{x}^{0}$. This
choice yields the maximum defect energy $\Delta E(L)$. Any other
choice $0 <\epsilon\leq\pi$ yields the same spin stiffness exponent
$\theta_{s}^{BT}$ defined by
\begin{equation}
\langle \Delta E_{s}^{BT}(L,\epsilon) \rangle =
{\cal A}(\epsilon) L^{\theta_{s}^{BT}}
\label{eq:fss.spin}
\end{equation}
The size $\epsilon$ of the twist from the BT value $\Delta^{0}_{x}$
affects only the amplitude ${\cal A}(\epsilon)$ which is a maximum
at $\epsilon =\pi$.
\newline\indent
A chiral domain wall is induced by imposing reflective BC\cite{sg-dw-kt}
which means that corresponding sites $(ij)$ on opposite faces are connected
by interactions $\tilde{V} = V(\theta_{i} + \theta_{j} - A_{ij})$ which is
equivalent to a reflection of the spins about some axis. In principle, one
follows the procedure for a spin domain wall to obtain the chiral defect
energy $\Delta E_{c} = E_{cD} - E_{0}$ where $E_{cD}$ is the minimum energy
of the system with these modified interactions $\tilde{V}$ connecting
opposite faces.
However, there is no reason to expect $E_{cD}>E_{0}$ as the BC defining
$E_{0}$ may trap a chiral defect in some samples and, in such cases, the
modified interactions $\tilde{V}$ will cancel the chiral defect to give
$E_{cD}<E_{0}$. This phenomenon has been observed previously in numerical
simulations of the $XY$ spin glass\cite{sg-dw-kt,sg-dw-mg}.
We therefore define the chiral
defect energy as $\Delta E_{c}^{BT}(L) \equiv |E_{cD}(L)-E_{0}(L)|$
and the chiral stiffness exponent $\theta_{c}^{BT}$ by a finite size
scaling ansatz analogous to Eq. (\ref{eq:fss.spin})
\begin{equation}
\label{eq:fss.chiral}
\langle \Delta E_{c}^{BT}(L) \rangle \sim L^{\theta_{c}^{BT}}
\end{equation}
Note that the Hamiltonian of Eq. (\ref{eq:h}) is truly invariant under
reflection
$\theta_{i}\rightarrow -\theta_{i}$ in the $XY$ spin glass case when
$A_{ij}=0,\pi$ as $A_{ij}=\pm\pi$ are equivalent.
The Hamiltonian with uniform distribution of $A_{ij}$ such as
a gauge glass is not truly invariant under reflection which would also require
$A_{ij}\rightarrow -A_{ij}$ for the Hamiltonian to be invariant
as it lacks the reflection symmetry.
\newline\indent
However, in an $XY$ spin glass there are two possible types of order,
spin glass order and
a chiral glass order each with their own stiffness exponent of
Eq. (\ref{eq:fss.spin}) and Eq. (\ref{eq:fss.chiral}). Recently, an important
prediction was made that $\theta_{s} = \theta_{c} <0$ for an $XY$ spin
glass in dimension $d<d_{l}$ where $d_{l}$ is the lower critical
dimension\cite{nh}. Although not rigorous, the arguments are very plausible
and supported by
analytic calculations on simple one dimensional systems in which
$\theta_{s} = \theta_{c}$ exactly\cite{nhm,tnh}. There is a notable lack of
analytic results in this field with which to test numerical simulations
and to our knowledge this is the {\it only} one existing at present.
We have checked our numerical method in $d=2<d_{l}$ and get agreement with
the analytic prediction that $\theta_{s} = \theta_{c} = -0.37\pm 0.015$
to within numerical uncertainty\cite{xysg-ka}.
Assuming the conjecture \cite{nh} is correct, this agreement
gives some confidence in our definition of
domain wall energies as discussed above and in our numerical method in
$d=3$ using the CG representation. There is no analogous equality of the
stiffness exponents in a $XY$ spin glass for $d =3>d_{l}$
so we do not attempt to estimate $\theta_{c}$ in $3D$ but concentrate
on the spin stiffness exponent $\theta_{s}$.
Also, at present, we are unable to derive an expression for
a CG Hamiltonian with reflective boundary condition in $d=3$.
All previous work on the $XY$ gauge glass\cite{gg-rtyf,gg-g,gg-ks,gg-mg}
and on the $XY$ spin glass\cite{sg-dw-kt,sg-dw-k,sg-dw-mg}
using the $T=0$ DWRG method have used different definitions for domain
wall energies. Minimization with respect to the global twists $\Delta_{\mu}$
is omitted, the lowest energy with $\Delta_{\mu} = 0$ is called $E_{p}$ and
the lowest energy with $\Delta_{\mu} = \pi$ is called $E_{ap}$. Neither of
these BC is compatible with the GS configuration as both must induce some
excitation from $E_{0}$. Nevertheless, the spin defect energy is defined by
$\Delta E_{s}^{RT} \equiv |E_{ap}-E_{p}|$ and the spin stiffness exponent
$\theta_{s}^{RT}$ by
\begin{equation}
\label{eq:rt}
\langle \Delta E_{s}^{RT}(L) \rangle  \sim L^{\theta_{s}^{RT}}
\end{equation}
We call this a random twist (RT) measurement as, for a particular sample,
the twists $\Delta_{\mu} = 0,\pi$ are two arbitrary random choices relative
to the best twist $\Delta_{\mu}^{0}$, which is the twist which yields the
lowest energy. In a uniform ferromagnet, $\Delta_{\mu}^{0}=0$ which is realized
by periodic BC and $\Delta_{\mu}^{0} + \pi$  by antiperiodic BC.
For a particular realization of randomness, $\Delta_{\mu}^{0}$ is the analogue
of periodic BC in a uniform ferromagnet.

\section{\bf Transformation to Coulomb Gas Representation}
\label{sec:cg}

\begin{figure}[tbp]
\center
\begin{minipage}{3.0in}
\epsfxsize= 3.0in \epsfysize=3.0in
\epsfbox{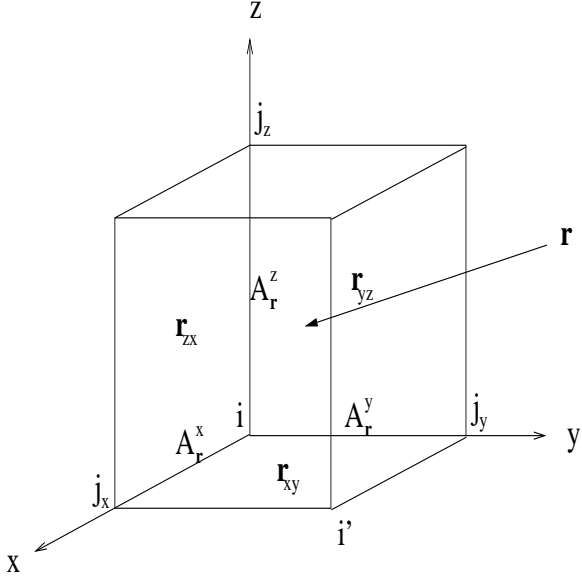}
\end{minipage}
\caption{The relation between the coordinate of the center of cube ${\bf r}$
(dual lattice site) and the coordinates of the plaquettes,
${\bf r}_{xy}$, ${\bf r}_{yz}$, and ${\bf r}_{zx}$,
associated with the cube at ${\bf r}$.
$i$ and $j_{\mu}$ are
sites of the original lattice.
The random bond variable $A_{ij_{x}}$ is relabelled by
$A_{\bf r}^{x}$ and similarly for the others. We assign three independent
random bond variables, $A_{\bf r}^{\mu}$ with $\mu=$ $x$, $y$, $z$,
to each cube at ${\bf r}$ as shown.}
\label{fig:cube1}
\end{figure}

\begin{figure}[tbp]
\center
\begin{minipage}{3.0in}
\epsfxsize= 3.0in \epsfysize=3.0in
\epsfbox{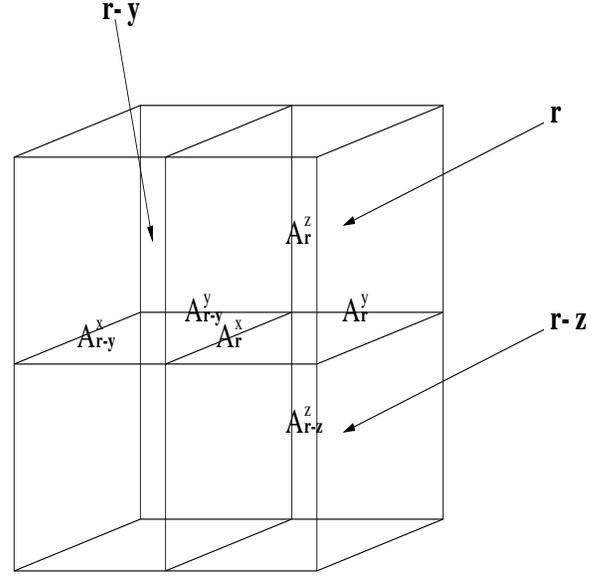}
\end{minipage}
\caption{Graphical explanation of our symbols. {\bf r} denotes
the center of the cube and light solid lines join original lattice sites.}
\label{fig:cube3}
\end{figure}

\begin{figure}[tbp]
\center
\begin{minipage}{3.0in}
\epsfxsize= 3.0in \epsfysize=3.0in
\epsfbox{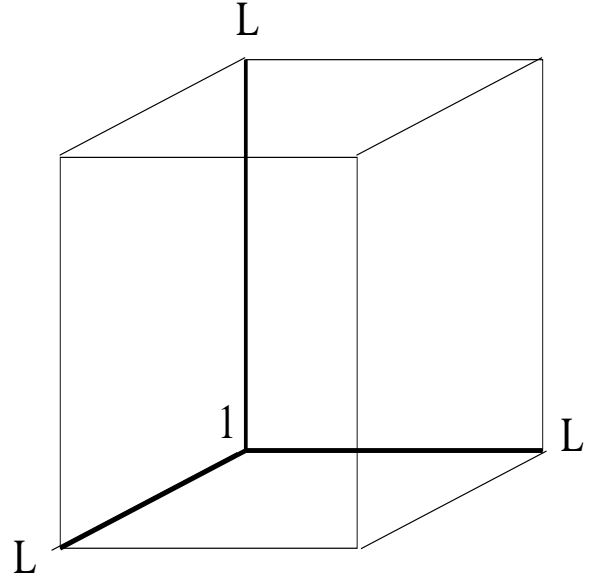}
\end{minipage}
\\
\caption{The $L \times L \times L$ system is represented by a cube.
The thick lines are our choices of the three global loops around the
whole system.}
\label{fig:cube2}
\end{figure}

In this section, we discuss the CG representation of the Hamiltonian of
Eq. (\ref{eq:h}) including all finite size contributions. This representation
parameterizes the energy in terms of the topological excitations on a torus
in $2D$ and a hypertorus in $3D$ and includes global excitations which
wind round the whole hypertorus. These latter excitations are very
important for a finite system and are vital for finite size scaling
considerations when one is limited to small system sizes $L$.
Also, every allowed configuration of
topological excitations is a local energy minimum \cite{kt,jmk}
as spin wave excitations decouple from the vortex excitations,
which allows us to obtain a more accurate
estimate of energy minima than using the phase representation of
Eq. (\ref{eq:h}) with the limited CPU time available. The transformation of
the two dimensional $XY$ model
to the CG representation including boundary contributions has been discussed
in detail in earlier works\cite{nh,duality1,duality2}. In this section, we
use the method of Ney-Nifle and Hilhorst\cite{nh} to transform to the CG
representation in $3D$.
\newline\indent
We first replace the potential $V(\phi)$ in Eq. (\ref{eq:h}) by
a piecewise parabolic potential which is equivalent
to a Villain\cite{villain} potential at $T=0$.
The partition function for a $L\times L\times L$ gauge glass model in $3D$ is
\begin{equation}
\label{eq:Z}
Z = \int_{-\pi}^{+\pi}\prod_{i}d\theta_{i}\sum_{\left\{n_{ij}\right\}}
\exp \Biggl[-\beta J \sum_{<ij>}
\Bigl(\theta_{ij} - A_{ij}\Bigr)^{2} \Biggr]
\end{equation}
where $\theta_{ij}\equiv\theta_{i} - \theta_{j} - 2\pi n_{ij}$ and
where $n_{ij}=-n_{ji}$ are integers on the bond $<ij>$.
By choosing one phase, $\theta_{0}$, as a reference,
the partition function can be written as
\begin{eqnarray}
&Z& = \int_{-\pi}^{\pi} d\theta_{0} \int_{-\infty}^{\infty}
\prod_{<ij>}d\theta_{ij}
 \exp \Biggl[ - \beta J \bigl( \theta_{ij} - A_{ij} \bigr)^{2}
 \Biggr] \nonumber \\
  & & \times\prod_{\bf r}
  \delta \Biggl( \sum_{P ({\bf r}_{xy}}) \theta_{ij} \bmod 2\pi
\Biggr)
 \delta \Biggl( \sum_{P ({\bf r}_{yz}}) \theta_{ij} \bmod 2\pi \Biggr) \cr
& &\times\prod_{\bf r}
 \delta \Biggl( \sum_{P ({\bf r}_{zx}}) \theta_{ij} \bmod 2\pi \Biggr)
 \delta \Biggl( \sum_{x} \theta_{ij} \bmod 2\pi \Biggr) \cr
& &\times\prod_{\bf r}
 \delta \Biggl( \sum_{y} \theta_{ij} \bmod 2\pi \Biggr)
 \delta \Biggl( \sum_{z} \theta_{ij} \bmod 2\pi \Biggr)
\end{eqnarray}
Here, ${\bf r}$ is the coordinate of the center of an elementary cube of the
original lattice which corresponds to the coordinate of a dual lattice site.
${\bf r}_{xy}$ is the coordinate of the center of the elementary plaquette in the
$xy$ plane and similarly for ${\bf r}_{yz}$ and ${\bf r}_{zx}$. Note that for a $2D$
system in the $xy$ plane, ${\bf r}_{xy}$ are the dual lattice sites ${\bf r}$.
Since each cube has six faces (plaquettes), each of which is shared by
two adjacent cubes, to each dual lattice site ${\bf r}$ we assign three
independent plaquettes with centers at ${\bf r}_{xy}$, ${\bf r}_{yz}$,
and ${\bf r}_{zx}$ as shown in Figs. (\ref{fig:cube1}) and (\ref{fig:cube3}).
$\sum_{P({\bf r}_{xy}}) \theta_{ij} $ is the circulation of $\theta_{ij}$
round the plaquette in the $xy$ plane of the cube at ${\bf r}$,
\begin{equation}
\sum_{P({\bf r}_{xy})} \theta_{ij} \equiv
\left[
  \theta_{ij_{x}}+\theta_{j_{x}i^{'}}-\theta_{j_{y}i^{'}}-\theta_{ij_{y}}
\right]
\end{equation}
and $\sum_{x} \theta_{ij}$ is the circulation of
$\theta_{ij}$ along an arbitrary loop in the $x$-direction round
the hypertorus,
\begin{equation}
\sum_{x} \theta_{ij} \equiv \sum_{i_{x}=1}^{L}
\theta_{(i_{x},i_{y},i_{z}),(i_{x}+1,i_{y},i_{z})}
\end{equation}
where $i_{\mu}$ with $\mu=x,y,z$ is the coordinate of the original lattice
site and $i_{y}$ and $i_{z}$ are fixed.
Other summations are defined similarly. Note that
one needs to consider only one global loop on the hypertorus in each
direction. Circulations round other global loops can be expressed in terms
of circulations round any three chosen global loops and
round elementary plaquettes. It is clear from the definition of $\theta_{ij}$
and periodic boundary condition imposed on the
$\theta_{i}$, these circulations  are integer multiples of $2\pi$.
Since the delta functions can be rewritten as follows,
\begin{eqnarray}
&\delta& \Biggl( \sum_{P({\bf r}_{xy})} \theta_{ij} \bmod 2\pi \Biggr)
=\frac{1}{2\pi}\sum_{n_{\bf r}^{z}=-\infty}^{\infty} \exp \Biggl[ i
n_{\bf r}^{z} \sum_{P({\bf r}_{xy})} \theta_{ij}
\Biggr] \nonumber\cr
&\delta& \Biggl( \sum_{x} \theta_{ij} \bmod  2\pi \Biggr)
= \frac{1}{2\pi} \sum_{n_{x}=-\infty}^{\infty}
\exp \Biggl[ in_{x} \sum_{x} \theta_{ij} \Biggr]
\label{eq:deltafn}
\end{eqnarray}
the partition function now becomes
\begin{eqnarray}
&Z& = Z_{0} \sum_{\bf n}
\sum_{ \{ {\bf n}_{\bf r} \} }
\int_{-\infty}^{\infty}\prod_{<ij>}d\theta_{ij}
\exp \Biggl[ -\beta J (\theta_{ij}-A_{ij})^{2} \Biggr] \cr
& \times &
\exp\Biggl[\sum_{\bf r}\Bigl(in_{\bf r}^{x}\sum_{P({\bf r}_{yz})}\theta_{ij}
 + in_{\bf r}^{y}
\sum_{P({\bf r}_{zx})}\theta_{ij} + in_{\bf r}^{z}\sum_{P({\bf r}_{xy})}
\theta_{ij}\Bigr)\Biggr] \cr
& \times & \exp \Biggl[ i n_{x} \sum_{x} \theta_{ij}+i n_{y} \sum_{y}
\theta_{ij}+i n_{z} \sum_{z} \theta_{ij} \Biggr]
\label{eq:z2}
\end{eqnarray}
where $n_{\bf r}^{\mu}$ and $n_{\mu}$ with $\mu = x,y,z$ are
integers, and $\sum_{\bf n} \equiv \prod_{\mu}\sum_{n_{\mu}}$
and
$\sum_{ \{ {\bf n}_{\bf r} \} } \equiv
\prod_{\mu}\sum_{n^{\mu}_{\bf r}}$
The sum $\sum_{\bf r}$ is over the dual
lattice sites ${\bf r}$ at the centers of the elementary cubes and $Z_{0}$ is
an unimportant constant. We use the notation $\sum_{P({\bf r}_{xy})}$ to denote
a sum over the bonds, in a clockwise direction, of the plaquette in the $xy$ plane
centered at ${\bf r}_{xy}$ and $\sum_{x}$ to denote a sum over bonds on a global
loop round the hypertorus in the $x$ direction.
To perform the integrations over $\{\theta_{ij}\}$, we choose the three global
loops around the hypertorus as shown in Fig. (\ref{fig:cube2}).
To deal with these global loops mathematically, we introduce the
following quantities,
\begin{equation}
\delta^{x}_{\bf r} \equiv \left \{ \begin{array}{ll}
     1    & \mbox {if ${\bf r}=(x,1,1)$ with $x=1,2, \cdots ,L$}  \\
     0    & \mbox {otherwise}
                      \end{array}
              \right.
\label{eq:deltaxr}
\end{equation}
similarly for $\delta^{y}_{\bf r}$ and $\delta^{z}_{\bf r}$. For example,
the cubes at ${\bf r}=$ $(x,1,1)$ have a part of the global loop
in the $x$ direction.
\newline\indent
The definition of $A_{\bf r}^{\mu}$ associated with the cube at ${\bf r}$
is also shown in Fig. (\ref{fig:cube1}). As the plaquettes,
one can assign three independent
$A_{\bf r}^{\mu}$ to each cube. After relabeling $A_{ij}$ by
$A_{\bf r}^{\mu}$, performing the integrations in Eq. (\ref{eq:z2})
over $\theta_{ij}$, the partition function becomes
\begin{eqnarray}
&Z& = Z_{0}\sum_{\bf n, n_{r}}
\exp \left[ -\frac{1}{4 \beta J}\sum_{\bf r} \sum_{\alpha}
\bigl\{ \left( {\bf \nabla} \times {\bf n}_{\bf r} \right)_{\alpha} +
        \delta^{\alpha}_{\bf r}n_{\alpha} \bigr\} ^{2} \right] \cr
& \times &
\exp \left[ -i \sum_{\bf r} \sum_{\alpha}
\bigl\{ \left( {\bf \nabla} \times {\bf n}_{\bf r} \right)_{\alpha} +
        \delta^{\alpha}_{\bf r}n_{\alpha}\bigr\} A^{\alpha}_{\bf r}
\right]
\label{eq:z3}
\end{eqnarray}
We use the following notation for the discrete derivative,
\begin{equation}
({\bf \nabla} \times {\bf n}_{\bf r})_{x}\equiv
(n_{\bf r}^{y}-n_{{\bf r}-\hat{{\bf z}}}^{y})
-(n_{\bf r}^{z}-n_{{\bf r}-\hat{{\bf y}}}^{z})
\end{equation}
and similarly for other components of ${\bf \nabla\times n_{r}}$.
Note that, when ${\bf r}=(1,y,z)$, then ${\bf r}-\hat{\bf x} =
(L,y,z)$ due to periodic boundary conditions, and similarly for ${\bf
r}-\hat{\bf y}$ and ${\bf r}-\hat{\bf z}$.
Applying the Poisson summation formula to $n_{\bf r}^{\mu}$,
\begin{equation}
\sum_{n=- \infty}^{\infty} f(n) = \sum_{q=- \infty}^{\infty}
\int_{- \infty}^{\infty} du e^{2 \pi iqu}f(u),
\label{eq:poisson}
\end{equation}
the partition function of Eq. (\ref{eq:z3}) becomes
\begin{eqnarray}
&Z & = Z_{0}\sum_{\bf n} \sum_{ \{ {\bf q}_{\bf r} \} }
\int_{-\infty}^{\infty} \prod_{\bf r} d{\bf u}_{\bf r}
\exp \Bigl[ 2\pi i\sum_{\bf r} {\bf q}_{\bf r} \cdot {\bf u}_{\bf r} \Bigr]
\nonumber \\
& \times &\exp \left[ -\frac{1}{4 \beta J}\sum_{\bf r} \sum_{\alpha=x,y,z}
\left\{ \left( {\bf \nabla} \times {\bf u}_{\bf r} \right)_{\alpha} +
        \delta^{\alpha}_{\bf r}n_{\alpha} \right\} ^{2} \right] \nonumber
\\
& \times &\exp \left[ -i \sum_{\bf r} \sum_{\alpha=x,y,z}
\left\{ \left( {\bf \nabla} \times {\bf u}_{\bf r} \right)_{\alpha} +
        \delta^{\alpha}_{\bf r}n_{\alpha}\right\} A^{\alpha}_{\bf r}
\right]
\label{eq:z4}
\end{eqnarray}
It is convenient to change integration variables to $v^{\mu}_{\bf r}$ to make
Eq. (\ref{eq:z4}) more symmetric
\begin{eqnarray}
v_{\bf r}^{x} &\equiv&
u_{\bf r}^{x}-\frac{y}{2L}n_{z}+\frac{z}{2L}n_{y}
\nonumber \\
v_{\bf r}^{y} &\equiv&
u_{\bf r}^{y}-\frac{z}{2L}n_{x}+\frac{x}{2L}n_{z}
          \\
v_{\bf r}^{z} &\equiv&
u_{\bf r}^{z}-\frac{x}{2L}n_{y}+\frac{y}{2L}n_{x}
\nonumber
\end{eqnarray}
where ${\bf r}=(x,y,z)$. The partition function of Eq. (\ref{eq:z4}) now becomes
\begin{eqnarray}
& Z &= Z_{0} \sum_{\bf n} \sum_{ \{ {\bf q}_{\bf r} \} }
\int_{-\infty}^{\infty} \prod_{\bf r} d{\bf v}_{\bf r}
 \exp \left[ 2\pi i \sum_{\bf r}
 {\bf q}_{\bf r} \cdot {\bf v}_{\bf r}
 \right] \nonumber \\
 & \times &\exp \left[ \frac{\pi i}{L} \sum_{\bf r}
 {\bf q}_{\bf r} \cdot \bigl( {\bf r} \times {\bf n}
 \Bigr\} \right] \nonumber \\
& \times & \exp \left[ -\frac{1}{4 \beta J}\sum_{\bf r} \sum_{\alpha=x,y,z}
\left\{ \left( {\bf \nabla} \times {\bf v}_{\bf r} \right)_{\alpha} +
        \frac{n_{\alpha}}{L} \right\} ^{2} \right] \nonumber \\
& \times &\exp \left[ -i \sum_{\bf r} \sum_{\alpha=x,y,z}
\left\{ \left( {\bf \nabla} \times {\bf v}_{\bf r} \right)_{\alpha} +
        \frac{n_{\alpha}}{L} \right\} A^{\alpha}_{\bf r} \right]
\label{eq:z5}
\end{eqnarray}
The terms linear in $n_{\mu}$ in the exponent of Eq. (\ref{eq:z5})
vanish due to the periodic boundary condition
$\sum_{\bf r} \left( v_{\bf r}^{x}-v_{{\bf r}-\hat{\bf y}}^{x} \right)=0$.
Eq. (\ref{eq:z5}) can be simplified by introducing frustration variables
$f_{\bf r}^{\mu}$ at site ${\bf r}$
\begin{eqnarray}
f_{\bf r}^{x} &\equiv& -\frac{1}{2\pi}\sum_{P({\bf r}_{yz})}A_{ij} \nonumber \\
&=&
\frac{1}{2\pi}\Bigl(A_{\bf r}^{y}-A_{{\bf r}+\hat{\bf z}}^{y} + A_{{\bf
r}+\hat{\bf y}}^{z} - A_{\bf r}^{z} \Bigr)
\label{eq:frustx}
\end{eqnarray}
from which $f_{\bf r}^{y}$ and $f_{\bf r}^{z}$ are obtained by cyclic
permutation of $xyz$. After some algebra, the partition function of
Eq. (\ref{eq:z5}) becomes a form suitable for integration over
$v_{\bf r}^{\mu}$
\begin{eqnarray}
Z &=& Z_{0} \sum_{\bf n} \sum_{ \{ {\bf q}_{\bf r} \} }
\int_{-\infty}^{\infty} \prod_{\bf r} d{\bf v}_{\bf r}^{x}
 \exp \left[ \frac{2\pi i}{2L} \Bigl\{
 {\bf q}_{\bf r} \cdot \bigl( {\bf r}\times{\bf n} \bigr)
                               \Bigr\} \right] \nonumber \\
  & &
\times \exp \left[ -\frac{1}{4 \beta J}\sum_{\bf r} \sum_{\alpha=x,y,z}
 \Bigl\{ \left( {\bf \nabla} \times {\bf v}_{\bf r} \right)_{\alpha}
  \Bigr\} ^{2} \right] \nonumber \\
  & &
\times \exp \left[ 2 \pi i \sum_{\bf r} ( {\bf q}_{\bf r}- {\bf f}_{\bf r} )
 \cdot {\bf v}_{\bf r} \right]
\label{eq:z6}
\end{eqnarray}
\newline\indent
To evaluate the integrals over $v_{\bf r}^{\mu}$ in Eq. (\ref{eq:z6})
it is convenient to take the Fourier transform
\begin{equation}
{\bf v}_{{\bf{r}}}=L^{- \frac{3}{2}} \sum_{{\bf{k}}}
e^{i {\bf{k}} \cdot {\bf{r}}} \tilde{{\bf v}}({\bf k})
\end{equation}
where ${\bf{k}}=(k_{x},k_{y},k_{z})$ and
$k_{i}=\frac{2\pi}{L} m_{i}$ with $m_{i}=0,1, \cdots ,L-1$.
We also decompose $\tilde{{\bf v}}({\bf k})$ into
longitudinal, $\tilde{{\bf v}}_{L}({\bf k})$, and transverse,
$\tilde{{\bf v}}_{T}({\bf k})$, components where
\begin{eqnarray}
\tilde{v}_{L}^{\alpha}({\bf k}) &   =     & \sum_{\beta}
\frac{(1-e^{ik_{\alpha}})(1-e^{-ik_{\beta}})}{\lambda_{\bf
k}}\tilde{v}^{\beta}({\bf k})
\nonumber \\
            & \equiv  & \sum_{\beta} L_{\alpha
\beta}\tilde{v}^{\beta}({\bf k})
\nonumber \\
\tilde{v}_{T}^{\alpha}({\bf k}) & \equiv  & \sum_{\beta} (\delta_{\alpha
\beta}-L_{\alpha \beta}) \tilde{v}^{\beta}({\bf k})
\nonumber \\
        & \equiv  & \sum_{\beta} T_{\alpha
          \beta}\tilde{v}^{\beta}({\bf k})
\nonumber \\
\lambda_{\bf k} &=& 6 - 2\cos k_{x} - 2\cos k_{y} - 2\cos k_{z}
\end{eqnarray}
define the longitudinal and transverse projection operators $L_{\alpha\beta}({\bf
k})$ and $T_{\alpha\beta}({\bf k})$.
In Fourier space, the integration over ${\bf v}_{\bf r}$ in
Eq. (\ref{eq:z6}) is straightforward
\begin{eqnarray}
\int \prod_{\bf k} d \tilde{\bf v}_{L} ({\bf k}) d \tilde{\bf v}_{T} ({\bf
k})
 \exp \left[ - \frac{1}{4 \beta J} \sum_{\bf k} \lambda_{\bf k}
 \mid \tilde{\bf v}_{T} ({\bf k}) \mid ^{2} \right] &   \nonumber \\
 \times \exp \left[ 2 \pi i \sum_{\bf k} \Bigl\{ \tilde{\bf p}_{T}({\bf k}) \cdot
 \tilde{\bf v}_{T}({\bf k}) + \tilde{{\bf p}}_{L}({\bf k}) \cdot
 \tilde{\bf v}_{L}({\bf k}) \Bigr\} \right]
\end{eqnarray}
where $\tilde{\bf p}({\bf k}) = {\bf q}({\bf k}) - {\bf f}({\bf k})$.
Integration over $\tilde{\bf v}_{L}({\bf k})$ gives $\tilde{\bf
p}_{L}({\bf k})= 0$. In real space, this is the condition that the discrete divergence of
the vorticity (charge) ${\bf q}_{\bf r}$ at any dual lattice site ${\bf r}$
obeys ${\bf \nabla}\cdot{\bf q_{r}} = 0$
since ${\bf\nabla}\cdot{\bf f_{r}} = 0$.
The integration over the transverse components
$\tilde{\bf v}_{T}({\bf k})$ are
simple Gaussian integrals and are easily performed. Integration over
${\bf\tilde{v}}_{L}(0)$ and ${\bf\tilde{v}}_{T}(0)$ yield the neutrality condition
\begin{equation}
\tilde{{\bf p}}(0)=\sum_{\bf r}{\bf p}_{\bf r}=0
\label{eq:neutral}
\end{equation}
The final step in this rather technical derivation of the Hamiltonian in
the CG representation is to apply the Poisson summation formula of
Eq. (\ref{eq:poisson}) to eliminate the $n_{\mu}$ in favor of
global vortices or charges $q_{\mu 1}$. The Gaussian
integrations yield the partition function
\begin{equation}
Z = Z_{0}\sum_{\{ {\bf q}_{\bf r} \}}\sum_{q_{\mu}}\exp\left[-\beta
H({\bf q}_{\bf r},{\bf f}_{\bf r},q_{\mu 1},f_{\mu 1})\right]
\label{eq:z7}
\end{equation}
The Hamiltonian $H$ is identified as
\begin{eqnarray}
H &=& (2\pi)^{2}J\sum_{{\bf r},{\bf r}'}({\bf q}_{\bf r} - {\bf f}_{\bf r})
\cdot ({\bf q}_{{\bf r}'} - {\bf f}_{{\bf r}'})G({\bf r}-{\bf r}')
\cr
 &+& \frac{J}{2L}\Bigl\{\frac{\pi}{L}\sum_{\bf r}\bigl( zp_{\bf r}^{y}
 - yp_{\bf r}^{z}\bigr) + Q_{x}\Bigr\}^{2}
  \cr
 &+& \frac{J}{2L}\Bigl\{\frac{\pi}{L}\sum_{\bf r}\bigl( xp_{\bf r}^{z}
 - zp_{\bf r}^{x} \bigr) + Q_{y}\Bigr\}^{2}
\cr
 &+& \frac{J}{2L}\Bigl\{\frac{\pi}{L}\sum_{\bf r}\bigl( yp_{\bf r}^{x}
 - xp_{\bf r}^{y} \bigr) + Q_{z}\Bigr\}^{2}
\label{eq:h3d}
\end{eqnarray}
where $G({\bf r}) = L^{-3}\sum_{{\bf k}\neq 0}(\exp (i{\bf k\cdot
r})-1)/\lambda_{\bf k} $ is the lattice Green's function on a simple
cubic lattice  in $3D$ with periodic BC and the quantities $Q_{\mu}$ are
\begin{equation}
Q_{x} = \pi\sum_{\bf r}(zp_{\bf r}^{y}\delta_{y,1} - yp_{\bf
r}^{z}\delta_{z,1}) + 2\pi L(q_{x1}-f_{x1})
\label{eq:Q}
\end{equation}
with $Q_{y}$ and $Q_{z}$ obtained from $Q_{x}$ by cyclic permutations of
$xyz$. In Eq. (\ref{eq:Q}), $f_{x1}$ is the circulation of $A_{ij}$ along
the chosen global loop round the hypertorus in the $x$ direction
\begin{equation}
2\pi f_{x1} = \sum_{x=1}^{L}A^{x}_{{\bf r}=(x,1,1)}
\label{eq:fx1}
\end{equation}
and similarly for $f_{y1}$ and $f_{z1}$. The integers $q_{\mu 1}$
are interpreted as circulations of the phase round the three independent global
loops encircling the hypertorus.
Eqs. (\ref{eq:h3d}) and (\ref{eq:Q}) are the main results of this section.
These expressions give the energy of a system of size $L$ on a hypertorus. To
find the minimum energy $E_{0}$ the Hamiltonian $H = H({\bf q}_{\bf
r},{\bf f}_{\bf r},q_{\mu 1},f_{\mu 1})$ is minimized with respect to the
bulk integer
valued vector charges ${\bf q}_{\bf r}$, the integer valued global winding
numbers $q_{\mu 1}$ and the global frustrations $f_{\mu 1}$. In the case
of the $XY$ spin glass the $q_{\mu 1}-f_{\mu 1}$ of Eq. (\ref{eq:Q}) are
restricted to
be integer or half integer while for the gauge glass this can have
any real value. Note that minimizing with respect to $q_{x1}$ and $f_{x1}$
is exactly minimizing with respect to the twist $\Delta_{x}$ and the best
twist $\Delta^{0}_{\mu}$ corresponds to $f^{0}_{\mu 1}$,
the value of $f_{\mu 1}$ at the energy minimum.
Adding a global twist $\Delta_{\mu}$ to the phases is
exactly equivalent to changing the global frustration from its original
value $f_{\mu 1}\rightarrow f_{\mu 1} + \Delta_{\mu}/2\pi$.
As discussed in Section \ref{sec:strategy}, a spin domain
wall is induced by $\Delta^{0}_{x} \rightarrow \Delta^{0}_{x} + \pi$
which, in the CG representation,
is $f^{0}_{x1} \rightarrow f^{0}_{x1} + 1/2$.
Note that the frustrations ${\bf f}_{\bf r}$ are given in terms of the random
$A_{ij}$ and are kept fixed during the minimization.
\newline\indent
To confront our numerical results with the {\it only} existing analytic
prediction\cite{nh} we need expressions for both spin and chiral domain
wall energies in dimension $d = 2 < d_{l}$. The CG representation of the $XY$
spin and gauge glasses including all finite size corrections
have been known for some time in $2D$\cite{duality2,nh} and,
for the sake of completeness, we quote the necessary results below.
To study the spin stiffness, we join
corresponding sites on opposite faces by
$\tilde{V} = V(\theta_{i} - \theta_{j} - A_{ij})$ and the CG Hamiltonian is
\begin{eqnarray}
H &=& (2 \pi)^{2} J \sum_{{\bf r}{\bf r}'} (q_{\bf r} - f_{\bf r})
G({\bf r} - {\bf r}') (q_{{\bf r}'} - f_{{\bf r}'})\cr
&+&\frac{J}{L^{2}}(\sigma_{x}^{2}
+\sigma_{y}^{2})
\label{eq:h2d}
\end{eqnarray}
where
\begin{eqnarray}
\label{def2d}
\sigma_{x}&=&-2\pi \biggl[ L(q_{x1}-f_{x1})
+ \sum_{\bf r} (q_{\bf r}-f_{\bf r}) y \biggr]
\cr
\sigma_{y}&=&-2\pi \biggl[ L(q_{y1}-f_{y1})
- \sum_{\bf r} (q_{\bf r}-f_{\bf r}) x \biggr]
\cr
G({\bf r})&=&\frac{1}{L^{2}} \sum_{{\bf k} \neq 0}
 \frac{e^{i {\bf k} \cdot{\bf r}}-1}{4-2 \cos k_{x} -2 \cos k_{y}}
\end{eqnarray}
Here, as in $3D$, ${\bf r}=(x,y)$ represents the coordinates of
the dual lattice sites and $G({\bf r})$ is the lattice Green's function
with periodic BC on the $\theta_{i}$. From Eq. (\ref{eq:Z}), we see that
the difference of a factor $2$ in the prefactor of Eq. (\ref{eq:h2d}) from
other works is in the coupling constant $J$. Also note that the
Hamiltonian of Eq. (\ref{eq:h2d}) describes the $XY$ spin glass when the
frustrations $f_{\bf r}$ and the global frustrations $f_{\mu 1}$ are
restricted to $(0,1/2)$. In the gauge glass, they can have any real value.
As in $3D$, the topological charges $q_{\bf r}$ are integers as are
$q_{\mu 1}$ because of the periodic BC in the $\theta_{i}$.
\newline\indent
The last piece of information we need is the CG representation of
the Hamiltonian of the $XY$ spin glass in $2D$ with a chiral domain
wall imposed. This is to be found in the paper of
Ney-Nifle and Hilhorst\cite{nh}.
A single chiral domain wall is induced by joining opposite
faces by interactions $\tilde{V} = V(\theta_{i}+\theta_{j}-A_{ij})$ which
is equivalent to imposing reflective BC\cite{sg-dw-kt,nh}. In turn, this is
equivalent to doubling the size in (say) the $x$ direction to a $2L\times L$
lattice in which the extra half is a charge conjugated image of the other.
This system has two chiral domain walls with Hamiltonian\cite{nh}
\begin{equation}
H_{R} = 2\pi^{2}J\sum_{{\bf r},{\bf r}'}(q_{\bf r}-f_{\bf
r})\tilde{G}({\bf r} - {\bf r}')(q_{{\bf r}'}-f_{{\bf r}'})
\label{eq:hcdw}
\end{equation}
where $\tilde{G}({\bf r})$ is the Green's function for a $2L\times L$
square lattice with {\it periodic} BC and also with
$q_{{\bf r}+L{\bf\hat x}}=
-q_{\bf r}$ and $f_{{\bf r}+L{\bf\hat x}} = -f_{\bf r}$.
Note that the sign reversal of the frustrations $f_{\bf r}$ is not
necessary for the spin glass
because $f_{\bf r}=\pm 1/2$ are equivalent.
\newline\indent
The form of the energy of Eq. (\ref{eq:hcdw}) is
used in the simulations to estimate $\theta_{c}$ the chiral stiffness exponent
as it is intuitively more transparent and more convenient than the
corresponding  expression with a single chiral domain wall\cite{nh}.
Unfortunately, we have been unable to derive the analogous expression in
$3D$ to Eq. (\ref{eq:hcdw}) so we have no independent estimate of $\theta_{c}$
\cite{sg-dw-mg} for the $3D$ $XY$ spin glass.
\section{\bf Numerical Method}
\label{sec:method}
\subsection{Minimization Algorithm}
\label{sec:algorithm}
In Section \ref{sec:strategy} we argued that it is necessary to find the
energies $E_{0}(L)$ and $E_{D}(L)$ essentially exactly for every sample to
control the errors in the {\em small} domain wall energy
$\langle \Delta E(L)\rangle$.
To our knowledge, for the systems of interest no algorithm exists
which can locate the global energy minima in polynomial time.
We are left with two methods:
(i) repeated simple quenches from an arbitrary initial configuration to $T=0$
 followed by a downward slide to the nearest local minimum and
(ii) simulated annealing\cite{anneal1} which is considerably more efficient
\cite{anneal2}.
By this we mean that, for the same CPU time, simulated annealing finds
a lower energy than simple quenching.
\newline\indent
We start with the system in some randomly chosen
configuration and quench to some $T_{0}$, determined by trial and error, for
each size $L$. Then we do a Monte Carlo (MC) sweep through the system by
inserting charges $q=1$ and $q'=-1$ on an arbitrary pair of
nearest neighbor sites of the
dual lattice in $2D$ and accepting or rejecting the move according to
conventional MC rules. This has the effect of inserting new charges,
annihilating charges or moving a charge by one lattice spacing
while maintaining charge neutrality $\sum q_{\bf r} = 0$.
In $3D$, the elementary excitation is a loop
of charge round an elementary square with vertices at dual lattice sites.
This maintains ${\bf\nabla\cdot q_{r}} = 0$.
The closed loop of charge can lie in
any of the three orthogonal planes of the cubic lattice and the charge can
circulate round the loop in either direction, making $6$ possibilities with
the center of the loop at a fixed but randomly selected position. The
temperature $T$ is then reduced to $T_{1}=\alpha T_{0}$ with $\alpha < 1$
whose value is again determined by trial and error and the procedure
iterated a large number, $N$, of times to reach a lowest temperature
$T_{N} = \alpha^{N}T_{0} \approx 0$. Of course, the system may be trapped in
a deep metastable well with barriers too high for the MC passes to overcome
so the whole annealing sequence is repeated $M$ times from different random
initial configurations and the lowest energy out of all the $NM$
trials recorded. Again, this does not guarantee that the global minimum
energy is found but this method does have a few checks built in. At the
crudest level, the best twist condition $E_{sD}(L) \geq E_{0}(L)$ must be
obeyed for each sample since $E_{0}(L)$ is, by construction, the lowest
energy of the system subject to the BC given by the interactions
$\tilde{V} = V(\theta_{i}-\theta_{j}-A_{ij})$ across opposite faces.
$E_{sD}(L)$ is obtained by $f^{0}_{x1}\rightarrow f^{0}_{x1}+1/2$ where
$f^{0}_{\mu 1}$ is the value of $f_{\mu 1}$ which makes the boundary terms of
Eqs. (\ref{eq:h3d}) and (\ref{eq:h2d}) vanish, corresponding to the best twist
$\Delta^{0}_{\mu}$. It is clear that $E_{sD}(L)$ is the energy of the system
containing an extra spin domain wall compared to the system with energy
$E_{0}(L)$. As discussed earlier, $E_{0}(L)$ is not necessarily the true GS energy
as the system may contain some chiral domain walls. However, by construction,
$E_{sD}(L)$ is the energy of the system with the {\it same} chiral defects
and an extra spin defect.
\newline\indent
If any sample violates the BT condition, clearly the annealing is not
sufficient and one can either increase the number $MN$ of annealing attempts or
just discard that sample. Increasing $MN$ involves a significant increase in
CPU time particularly for the larger sizes $L$ so the choice depends on the
time available. However, even if the BT condition is satisfied for every
sample, there is no guarantee  that the lowest energies found are {\it true}
global minima. To improve the chances that true minima are achieved, where
possible we did two independent simulations on identical samples with
different pseudo random number sequences and, if the same energy minima are
found in both simulations, this is defined to be the true minimum. This
procedure is very expensive in CPU time and our resources did not permit
this last check to be performed for every sample, particularly for the
largest systems. This check was done for at least a few randomly selected samples,
except for $L=7$ in the $3D$ gauge glass. Averaging over disorder was
performed over as many samples as possible with the aim of making the
uncertainty in the mean domain wall energy $\langle \Delta E(L) \rangle$
less than $3\%$ which requires averaging over at least $10^{3}$ samples. The
first set of checks are done on every sample which makes it fairly probable
that exact minima $E_{0}(L)$ and $E_{D}(L)$ are obtained. We can assume with some
confidence that the uncertainty in $\langle \Delta E(L) \rangle$ is purely
statistical and $O(N^{1/2})$. Averaging over about $10^{3}$ samples leads to an
acceptable uncertainty of about $3\%$ in $\langle \Delta E(L) \rangle$.
\newline\indent
\subsection{$XY$ Spin Glass in $2D$}
\label{sec:xysg2d}
\begin{figure}[tbp]
\center
\begin{minipage}{3.0in}
\epsfxsize 3.0in
\epsfbox{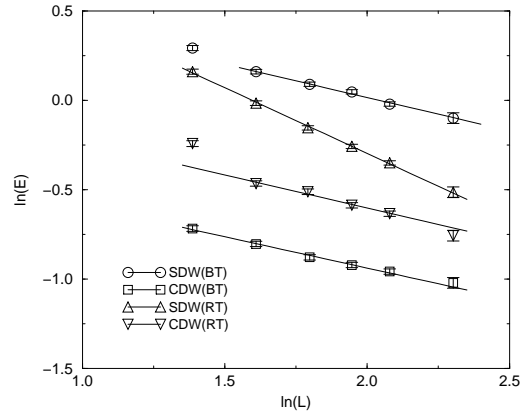}
\end{minipage}
\caption{Top to bottom: $L$ dependence of $\Delta E_{s}^{BT}$,
$\Delta E_{s}^{RT}$, $\Delta E_{c}^{RT}$, $\Delta E_{c}^{BT}$ for the $2DXY$ spin glass.}
\label{fig:xysg2d}
\end{figure}

This particular system with a Hamiltonian of Eq. (\ref{eq:h}) in $2D$ is not
particularly interesting in the sense that there exists no finite
temperature transition.
It has been known for a long time that $d_{l}>2$ and both spin and chiral
stiffness exponents are negative.
However, the situation has been controversial due to
the possible existence of different stiffness or correlation
length exponents for spin and chiral glass order in $2D$.
Previous estimates of the spin and chiral stiffness exponents are
summarized by $\theta_{s} \approx 2\theta_{c} \approx -0.78$
based on extensive numerical simulations such as DWRG and
finite temperature MC simulations \cite{sg-dw-kt,sg-dw-mg,sg-dw-k,sg-by,sg-rm}.
However, Ney-Nifle and Hilhorst\cite{nh} made a non-rigorous but very
plausible conjecture based on analytic considerations that for dimension
$d \leq d_{l}$, $\theta_{s} = \theta_{c} \leq 0$. This is supported by
exact analytic results on simple models (i) a $XY$ spin glass on a ladder lattice
\cite{nhm} where the common correlation length exponent $\nu = |\theta|^{-1}=0.5263...$
and (ii) the $XY$ spin glass on a tube lattice \cite{tnh} with $\nu = 0.5564...$.
The key disagreement between numerical estimates and analytic theory is in the conjectured
equality of $\theta_{s}$ and $\theta_{c}$. If one makes the hypothesis that earlier
estimates are in error, then the most likely reason is
that the simulations are computing the wrong quantity when estimating one or
both of the stiffness exponents. It is extremely unlikely that all the
simulations are in error for simple technical reasons as all find the same
values of $\theta_{s}$ and $\theta_{c}$, but $\theta_{s}\neq\theta_{c}$.
\newline\indent
To test this hypothesis,
we have performed simulations of the $2DXY$ spin glass in the CG
representation using Eq. (\ref{eq:h2d}) to estimate the spin stiffness exponent
$\theta_{s}$ using both BT and RT measurements with the results
$\theta^{BT}_{s} = -0.37\pm 0.015$ and
$\theta^{RT}_{s} = -0.76\pm 0.015$ \cite{xysg-ka}.
These numbers were obtained from sizes $L = 4,5,6,7,8,10$ averaging over
$2560$ samples for $L\leq 8$ and $1152$ samples for $L=10$. As expected,
the value of $\theta^{RT}_{s}$ agrees with all previous estimates
\cite{sg-dw-kt,sg-dw-mg},
all of which use the RT measurement in some form. Both the BT and the RT
data fit the scaling ansatz of Eq. (\ref{eq:fss}) equally well and some other
information is needed to decide which value of $\theta_{s}$, if either, is correct.
Both cannot be correct as both are supposed to measure the same quantity. The necessary
information is in the chiral stiffness exponent $\theta_{c}$ which we
measure by simulating Eq. (\ref{eq:hcdw}) on a $2L\times L$ system.
Again, both BT and RT
measurements were made with the same range of $L$ and the same number of
samples as for $\theta_{s}$ with the results
$\theta_{c}^{BT} = -0.37\pm 0.010$
and $\theta_{c}^{RT} = -0.37\pm 0.015$.\cite{xysg-ka}
At first sight it is surprising
that both measurements give the same value for $\theta_{c}$ to within numerical
uncertainty while the values of $\theta_{s}^{RT}$ and $\theta_{c}^{BT}$
differ by a factor of $2$. Note that the boundary terms $\sigma_{\alpha}$ of
Eq. (\ref{eq:h2d}) which contain the twist parameter $f_{\mu 1}$ vanish in
the BT condition and such boundary terms are absent from Eq. (\ref{eq:hcdw}).
Thus, any measurement with a CG representation in which boundary contributions
are absent is automatically a BT measurement.
Since $\theta_{c} = \theta_{s}^{BT} \approx -0.37$
and $\theta_{c} \neq \theta_{s}^{RT} \approx -0.76$,
{\it assuming} that the conjecture is correct,
we conclude that the BT measurement yields a reasonable estimate of the true
$\theta_{s}$ while the commonly used RT measurement yielding
$\theta_{s}^{RT} \neq \theta_{c}$ is not an appropriate method for the
small values of $L$ accessible at present.
Our simulation results are shown in Fig. (\ref{fig:xysg2d}).
We do not understand what, if anything,
$\theta_{s}^{RT}$ means despite the apparent excellent fit of
$\langle \Delta E_{s}^{RT}(L) \rangle$ to the scaling ansatz
as this is the energy difference
of the system subject to two random choices of BC and we can see no reason
why it should scale as $L^{\theta_{s}}$ over decades of $L$.
\section{RESULTS}
\label{sec:results}
Our investigation of the $2DXY$ spin glass establishes that the BT method
of measuring domain wall energies can reproduce the one and only available analytic
prediction \cite{nh} which is relevant for our purposes. This
gives some encouragement to venture into areas where no such analytic guide
exists. In this Section, we report some new results on the spin stiffness
exponent $\theta_{s}$ for the $XY$ spin glass in $3D$ (\ref{sec:3dxysg}),
and the gauge glass in both $2D$ and $3D$ (\ref{sec:gg}).
We also perform simulations on systems with varying strengths
of disorder (\ref{sec:xyrdm}) where we study the
renormalization group flows for both the coupling constants $J(L)$
and disorder strength $A(L)$ with increasing length scale $L$.
All our results are numerical and the exponent
$\theta_{s} = \theta_{s}^{BT}$ is estimated by fitting estimates of
$\langle \Delta E_{s}^{BT}(L) \rangle$ to the scaling ansatz of
Eq. (\ref{eq:fss}). The system
sizes $L$ are very small as they are severely constrained by the necessity
of controlling the errors in both $E_{0}(L)$ and $E_{D}(L)$, which are
estimated by independent simulations.
However, the fit to the scaling ansatz is
very good despite the very few data points for all cases we have
studied and is comparable with the same procedure carried out on the $2D$ Ising
ferromagnet. In this case, simulations for sizes $L=2,3,4,5$ are
sufficient to reproduce the exact $\theta_{s} = 1$ to high accuracy.
\subsection{$XY$ spin glass in $3D$}
\label{sec:3dxysg}
\begin{figure}[tbp]
\center
\begin{minipage}{3in}
\epsfxsize= 3in \epsfbox{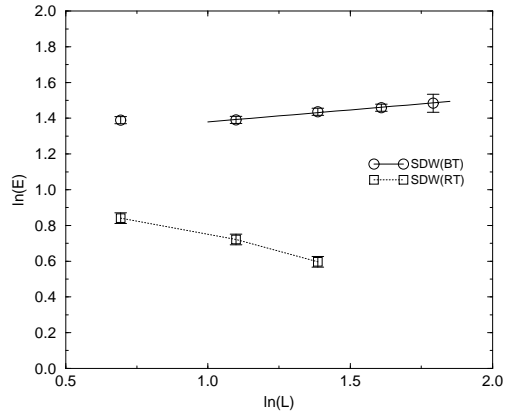}
\end{minipage}
\\
\caption{$L$ dependence of $\Delta E_{s}^{BT}$ and
$\Delta E_{s}^{RT}$ in $3D$. The error in the $L=6$ point is due to
rather few samples. Solid line is power law fit and the dotted line is
a guide to the eye.}
\label{fig:xysg3d}
\end{figure}

This system has been somewhat controversial for some years and is not yet
settled. It has been believed that
that $d_{l} \geq 4$ for spin glass order, and $d_{l}<3$ for chiral
glass order and that, in $2D$ and $3D$, spin and chiral variables decouple
and order separately \cite{sg-mc-kt,sg-dw-kt,sg-dw-k,sg-dw-mg,sg-by,sg-rm}.
This allows for the widely accepted scenario that in $3D$ spin glass order
sets in at $T_{SG}=0$ whereas chiral glass order sets in at $T_{CG} > 0$.
This scenario is based on MC simulations at finite $T$\cite{sg-dw-k,sg-jy}
and on $T=0$ defect energy
scaling \cite{sg-dw-kt,sg-dw-k} using the RT method.
Attempts to show rigorously that
$T_{SG} =0$ in $3D$\cite{ana-no} fail if reflection symmetry is
broken\cite{sg-dw-k,ana-sy,ana-on} at finite $T$.
The first cracks in this widely
accepted scenario appeared recently when Maucourt and Grempel\cite{sg-dw-mg}
published the results of a large scale defect energy scaling study of the
$3D$ $XY$ spin glass model of Eq. (\ref{eq:h}). They used the RT method, as all
previous defect energy scaling studies have done, with sizes $L \leq 12$ in
$2D$ and $L \leq 8$ in $3D$.
Although the fit to the scaling ansatz is not good due to strong
crossover effects and large uncertainties in the large $L$ data
which was used to estimate $\theta_{s,c}^{RT}$,it is clear that both
$\langle \Delta E_{s}^{RT}(L) \rangle$and $\langle \Delta E_{c}^{RT}(L)\rangle$
are increasing with $L$ which implies that there
is both spin and chiral glass order at sufficiently small $T >0$. However,
as argued above a valid numerical method must yield $\theta_{s} = \theta_{c}$
in $2D$ while they obtain $\theta_{s} \approx 2\theta_{c} \approx
-0.78$\cite{sg-dw-mg} in agreement with other estimates \cite{sg-dw-kt,sg-dw-k}.
Finite $T$ simulations seem to suffer from severe equilibration difficulties
\cite{sg-dw-k,sg-jy} which makes any conclusions from them also suspect.
\newline\indent
In view of the lack of reliable results for the $3DXY$ spin glass, we have
done some preliminary simulations on very small systems with $L = 2,3,4,5,6$
in the CG representation with Eq. (\ref{eq:h3d}) where $f_{\mu 1} =0,1/2$ in
Eq. (\ref{eq:Q}). Following the method outlined in Section \ref{sec:xysg2d}
we estimate $\theta_{s}^{BT} = +0.10\pm 0.03$ while the data for
$\langle\Delta E_{s}^{RT}(L)\rangle$ is clearly decreasing with $L$ for
$L=2,3,4$, with roughly the same slope as found by Kawamura \cite{sg-dw-k}
for the same range of $L$.
Our value $\theta_{s}=0.10\pm 0.03$ \cite{xysg-ka} is to be compared with the recent
estimate $\theta_{s}^{RT} = +0.052 \pm 0.03$ for $L =5,6,7$ \cite{sg-dw-mg}. Although
these values are close numerically and are equal to within error bars, we are
more inclined to believe in the former number as we consider $\theta_{s}^{BT}$
to give the true spin stiffness exponent and we speculate that further
points will also lie on the scaling ansatz and will significantly
reduce the $30\%$ uncertainty. The results are shown in Fig. (\ref{fig:xysg3d}).
Unfortunately, we have
no estimate for the chiral stiffness exponent $\theta_{c}^{BT}$ for the $3DXY$
spin glass as we have been unable to derive the $3D$ analogue of
Eq. (\ref{eq:hcdw}) to estimate $\Delta E_{c}^{BT}(L)$ and
$\Delta E_{c}^{RT}(L)$. However, because the numerical values of
$\theta_{c}^{BT}$ and $\theta_{c}^{RT}$ are the same in $2D$ and equal to
$\theta_{s}$ as required, we are reasonably safe in assuming that existing
estimates\cite{sg-dw-k,sg-dw-mg} of $\theta_{c}$ in $3D$ are fairly accurate.
Even though our spin stiffness exponent $\theta_{s}^{BT} \approx +0.1$
has a rather large uncertainty, it suggests that the lower critical
dimension for spin glass order is $d_{l}<3$, as for chiral order.
This is to be expected from analytic arguments \cite{nh}.
\subsection{Gauge Glass in $2D$ and $3D$}
\label{sec:gg}
\begin{figure}[tbp]
\center
\begin{minipage}{3.0in}
\epsfxsize=3.0in  \epsfysize=3.0in
\epsfbox{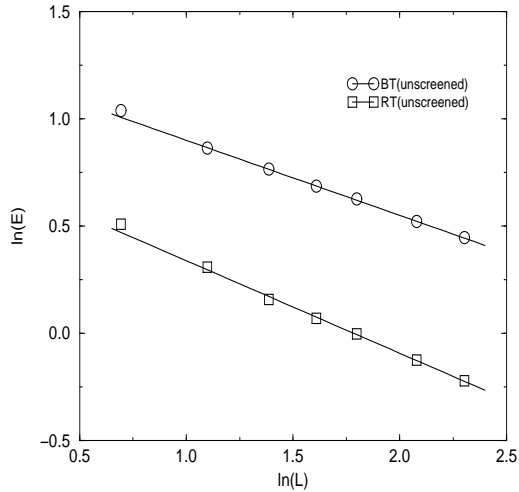}
\end{minipage}
\caption{Size $L$ dependence of domain wall energy in $2D$. Both RT
  and BT measurements are shown. Solid lines are power law fits.
Error bars are not shown if smaller than symbol size.}
\label{fig:gg2d}
\end{figure}

\begin{figure}[tbp]
\center
\begin{minipage}{3.0in}
\epsfxsize= 3.0in  \epsfysize=3.0in
\epsfbox{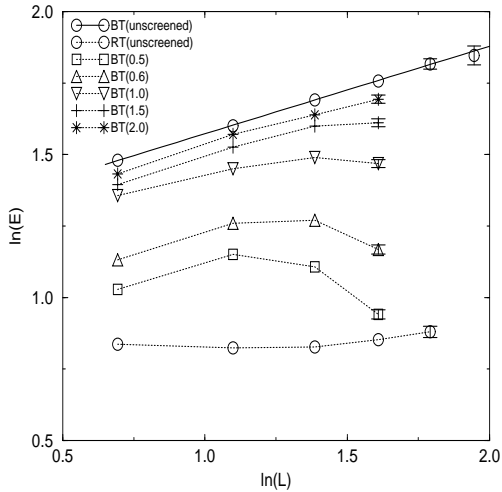}
\end{minipage}
\caption{$L$ dependence of domain wall energy in $3D$. Bottom curve is
RT measurement for unscreened interaction.
All others are BT measurements. Topmost curve is unscreened case
$L=2-7$. Other curves are screened interactions with $\lambda$
decreasing from top to bottom. Solid line is a power law fit and
dotted lines are guides for the eye.}
\label{fig:gg3d}
\end{figure}

We have also performed simulations on the gauge glass in the CG
representation using Eq. (\ref{eq:h2d}) in $2D$ and Eq. (\ref{eq:h3d}) in $3D$.
The only differences to the spin glass are in the values of $f_{\mu 1}$ and
$f_{\bf r}$ which can have any value in the interval $[-1/2,1/2)$. The frustrations
$f_{\bf r} = -\sum_{P({\bf r})}A_{ij}/2\pi$ are correlated random
variables as the $A_{ij}$ are the independent random variables. Similarly, in $3D$,
the frustrations ${\bf f_{r}}$ of Eq. (\ref{eq:frustx}) are correlated random
variables with each component in the interval $[-1/2,1/2)$. The three global frustrations
$f_{\mu 1}$ can take any value in the same interval. In both spin and gauge glasses, the
vorticities $q^{\mu}_{\bf r}$ and $q_{\mu 1}$ are integers. We have not computed
$\theta_{c}$ in either $2D$ or $3D$ for the gauge glass, mainly because we
have been unable to obtain the appropriate expression for the Hamiltonian
in the CG representation with reflective BC in $3D$ and we are unable to
understand what information such a simulation would yield. The major
reason for doing such a simulation in the $2D$ spin glass case was to confirm that
our procedure is a useful numerical method to estimate the actual spin stiffness exponent
$\theta_{s}$ for both spin and gauge glasses in $2D$ and $3D$.
To compare with earlier work \cite{gg-rtyf,gg-g,gg-ks,gg-mg} we have also estimated
$\theta_{s}^{RT}$ by keeping $f_{\mu 1}$ fixed at some random value during the
minimization, exactly as for the spin glass. The results of the simulations \cite{gg-ka}
in $2D$ for system sizes $L = 2,3,4,5,6,8,10$ are shown in Fig. (\ref{fig:gg2d}) for
$\langle\Delta E_{s}^{BT}(L)\rangle$ and $\langle\Delta E_{s}^{RT}(L)\rangle$.
Averages were performed over about $10^{3}$
samples for each size $L$. In this case, all the checks discussed in section
\ref{sec:algorithm} were performed so we assume the errors are purely
 statistical. Both fit well to the scaling ansatz of
Eq. (\ref{eq:fss}) with very similar errors with the values
$\theta_{s}^{BT} = -0.36\pm 0.010$ and $\theta_{s}^{RT} = -0.45\pm 0.015$.
The latter value is consistent with all earlier estimates of $\theta_{s}$
\cite{gg-rtyf,gg-g,gg-mg}
which is not a surprise as all these were done using the RT method in some
form. However, as argued in section \ref{sec:xysg2d}, this is not an accurate
estimate of $\theta_{s}$ so this number is {\em not} the $T=0$ spin stiffness
exponent. A more accurate estimate of this is
$\theta_{s}^{BT} = -0.36\pm 0.010$ which is significantly larger than
$\theta_{s}^{RT}$, so that the $2D$ gauge glass has a longer correlation
length $\xi (T)\sim T^{-1/|\theta_{s}|}$ than previously thought.
\newline\indent
We have also obtained some estimates of $\theta_{s}$ in $3D$ by performing
simulations\cite{gg-ka} of the gauge glass in the CG representation using
Eq. (\ref{eq:h3d}) with the distribution of frustrations ${\bf f}_{\bf r}$
appropriate for this case as determined by taking the $A_{ij}$ uniformly
distributed in $(-\pi ,+\pi ]$. The system sizes are $2\leq L\leq 7$ with
disorder averaging over $10^{3}$ samples for $L\leq 5$, $300$ for $L =6$
and $60$ for the largest size $L = 7$. The uncertainty in $\langle\Delta
E_{s}^{BT}(L)\rangle$
for $L=7$ is very large, but this data point is included to check that it is
consistent with the behavior deduced from the more reliable data of the
smaller sizes. The results are shown in Fig. (\ref{fig:gg3d})
for $\langle\Delta E_{s}(L)\rangle$
for the unscreened gauge glass using both BT and RT measurements and for
a gauge glass in $3D$ with screened interactions, $\lambda <\infty$.
The BT data fit the scaling
form of Eq. (\ref{eq:fss}) very well for sizes $L\leq 6$  with exponent
$\theta_{s}^{BT} = +0.31\pm 0.010$. If the $L=7$ point is also included in
the fit, we obtain $\theta_{s}^{BT} = +0.30\pm 0.015$. These errors in
$\theta_{s}$ come from a naive least squares fit of the data to a straight
line and should not be taken too seriously. The $L=7$ data is suspect because
$4$ samples violated the BT condition $\Delta E_{s}^{BT}(L) \ge 0$ out of a
total of only $64$,
despite running highly vectorised code for a few thousand CPU
hours on a Cray J90. Time did not permit any further checks for
attaining the global energy minimum for $L=7$. We cannot be sure that the
remaining $60$ samples which did not violate the BT condition reached their
global energy minima nor that the energies $E_{sD}$ are determined
sufficiently accurately.
The error bar on the $L=7$ point in Fig. (\ref{fig:gg3d}) assumes that the uncertainty
in $\langle\Delta E_{s}^{BT}(L=7)\rangle$ is purely statistical
and the true uncertainty is probably {\em much} larger.
At least an order of magnitude more CPU time is needed
for sufficient annealing to reach the true minima and to perform the
additional simulations with different random number sequences to check
that the minimization algorithm is successful.
This is just for a single batch of
$64$ samples and to reduce the uncertainty to $3\%$, yet another order of
magnitude of CPU time would be needed to average over $10^{3}$ samples.
This is totally out of reach with the computing resources available to us.
What data we have is entirely consistent with the scaling form of
Eq. (\ref{eq:fss}) with $\theta_{s}\approx +0.30$ with no sign of
any deviation from this form.
\newline\indent
The behavior of $\langle\Delta E_{s}^{RT}(L)\rangle$ is also shown in
Fig. (\ref{fig:gg3d}) for sizes
$L\leq 6$ which is very much like the data obtained by earlier
simulations.
This clearly does not fit the scaling form of Eq. (\ref{eq:fss}) for these
small values of $L$, but if one insists on extracting a value of
$\theta_{s}^{RT}$ from the data, one obtains consistency with previous estimates
\cite{gg-g} for the spin stiffness exponent
$\theta_{s}^{RT} \approx +0.05\pm 0.05$. As can be seen from
Fig. (\ref{fig:gg3d}), this estimate has no
meaning as the data
clearly does not obey the scaling ansatz. In fact, as
noticed by Maucourt and Grempel\cite{gg-mg},
$\langle\Delta E_{s}^{RT}(L)\rangle$ seems to
start increasing with $L$ for $L >5$ but, as we argue earlier,
this may, or may not, eventually scale as $\langle\Delta
E_{s}^{BT}(L)\rangle$ for sufficiently large $L$.
Speculation along these lines is fruitless until computers which are many
orders of magnitude faster become available or until an analytic solution
is found. We conclude that, for the {\em unscreened} gauge glass in $3D$, the
spin stiffness exponent $\theta_{s} = +0.31\pm 0.010$, which is considerably
larger than earlier estimates and indicates that the lower critical
dimension $d_{l}< 3$. This is consistent with finite $T$ MC results
\cite{gg-hs,gg-rtyf,gg-wy} for the gauge glass in $3D$ which indicate that
$T_{c} ={\cal O}(J)$. This value of $T_{c}$ is very difficult to reconcile with
the suggestion that $d_{l}\approx 3$ from previous DWRG studies \cite{gg-rtyf,gg-g,gg-ks,gg-mg}
as this implies a very small value of $T_{c}/J$.
\newline\indent
We have also studied the effects of screened vortex-vortex interactions
on the spin domain wall energy using the BT measurement in $3D$.
Screening of the Coulomb interaction of
vortices is implemented by adding a mass term $\lambda^{-2}$ to the
denominator of the Green's function \cite{gg-by}
\begin{equation}
G({\bf r}) = \frac{1}{L^{3}}\sum_{{\bf k}\neq 0}\frac{e^{i{\bf k\cdot r}}
- 1}
{6-2{\rm cos}k_{x}-2{\rm cos}k_{y}-2{\rm cos}k_{z}+\lambda^{-2}}
\label{eq:greenscreen}
\end{equation}
The results are also shown in Fig. (\ref{fig:gg3d}). We average over $10^{3}$
samples for $L = 2,3,4$ and $250$ for $L=5$ for several values of the screening
length $\lambda$. Screening is clearly a relevant perturbation when $\lambda$
is finite and $\theta_{s}^{BT} <0$ but our small sizes do
not permit an estimate of the value of $\theta_{s}^{BT}$. For large screening
lengths, $\langle\Delta E_{s}^{BT}(L)\rangle$ seems
to scale the same as for the unscreened case but we expect that
$\langle\Delta E_{s}^{BT}(L)\rangle$ will
decrease as $L^{\theta_{s}}$ with $\theta_{s} < 0$ at length scales which
are beyond our computing power for any $\lambda <\infty$.
These results are consistent with those of Bokil and Young\cite{gg-by}
who studied the question of screening using a RT measurement and with
Kisker and Rieger \cite{kisker} for very strong screening.
\newline\indent
\subsection{Varying disorder strength in $2D$ and $3D$}
\label{sec:xyrdm}
\begin{figure}[tbp!]
\center
\begin{minipage}{3.0in}
\epsfxsize= 3.0in \epsfysize=3.0in
\epsfbox{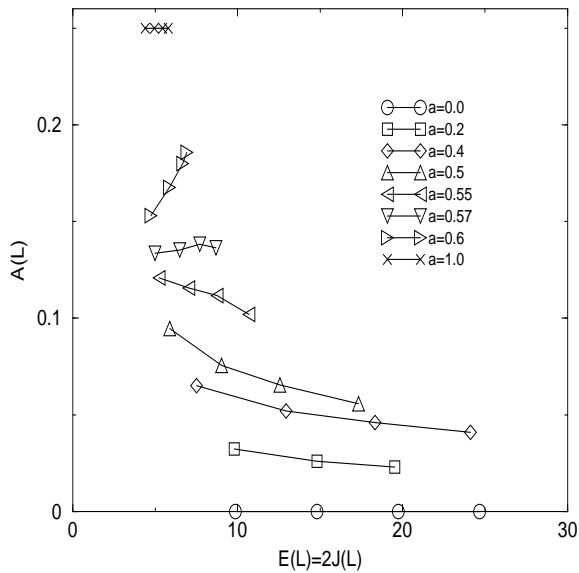}
\end{minipage}
\caption{RG flow in $3D$. The flows are from left to right for all $\alpha$.}
\label{fig:rgflow3d}
\end{figure}

We have also performed simulations with various strengths of disorder
in the CG representation
using Eq. (\ref{eq:h2d}) for the $2D$ case and Eq. (\ref{eq:h3d}) for $3D$
where the random bond variables $A_{ij}$
are independently uniformly distributed in the range
$[-\alpha\pi,\alpha\pi)$ with $0\le\alpha\le 1$ so that
$\langle A_{ij} \rangle =0$ and $\langle |A_{ij}| \rangle=\alpha\pi/2$.
Physical realizations of this model are, e.g., an $XY$ magnet with
random Dzyaloshinski-Moriya interactions \cite{xy-magnet} and
Josephson junction arrays with positional disorder \cite{gk1,gk2}
where both the effective coupling constant $J(L)$ and the
effective disorder strength $A(L)$ at length scale $L$ play a role.
Studies in $2D$ \cite{xy-magnet,gk1,gk2} suggest that
weak disorder $(\alpha\agt 0)$ does not affect the existence of an
ordered phase at intermediate temperature but there is a
re-entrant transition to a disordered phase at low temperature.
However, recent analytic \cite{xyrdm-nsk,xyrdm-kn,xyrdm-s,xyrdm-cf} and
numerical \cite{gg-ks} studies show there is an ordered phase for
$T<T_{c}(\alpha)$ with $T_{c}(\alpha)>0$ for $0\le\alpha<\alpha_{c}$.
\begin{figure}[tbp]
\center
\begin{minipage}{3.0in}
\epsfxsize=3.0in \epsfysize=3.0in
\epsfbox{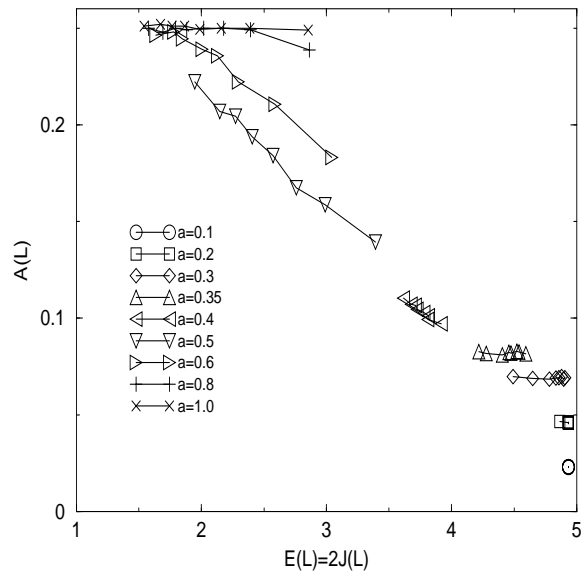}
\end{minipage}
\caption{RG flow in $2D$. The flows are from right to left
for $\alpha \ge 0.4$ and from left to right for $\alpha \le 0.35$. }
\label{fig:rgflow2d}
\end{figure}
To study how the disorder strength behaves as the length scale $L$
varies, one must, if possible, identify the scaled disorder strength
$A(L)$ with some measurable quantity.
An identification has been proposed in a recent numerical study
\cite{gg-ks} where the effective disorder strength is defined as
$2\pi A(L) \equiv \langle | {\bf \Delta}^{0}(L)| \rangle$ with
$A(1)=\alpha/4$ so that one can follow the flows of both $J(L)$ and $A(L)$
with increasing length scale $L$. With this definition, $0\leq A(L)\leq 1/4$.
${\bf \Delta}^{0}(L)$ is the global phase twist minimizing
the energy of a system of size $L$ for a particular realization of
disorder. For two phases with energy
$E_{12} = V(\theta_{1} - \theta_{2} - A_{12})$, the minimum is at $\theta_{1} - \theta_{2}
= A_{12}$ which is satisfied by applying a ``global'' phase twist of $A_{12}$.
Hence, follows the definition of $A(L)$ as a measure of disorder at scale $L$.
\newline\indent
Since the numerical study \cite{gg-ks} used the Hamiltonian of Eq. (\ref{eq:h})
in the phase representation, we re-investigate this model in the CG representation.
The simulations were performed for $L=2,3,4,5,6,8,10$ in $2D$,
$L=2,3,4,5$ in $3D$ and averaged over at least $10^{3}$ samples.
The results are shown in Fig. (\ref{fig:rgflow2d}) in $2D$
and Fig. (\ref{fig:rgflow3d}) in $3D$.
We are interested in the stable fixed point values at $L\rightarrow\infty$
$J^{*}$ and $A^{*}$ as these determine the nature of the phases.
In $2D$, weak disorder $(\alpha < \alpha_{c} \approx 0.37)$ seems
to be marginal and the system seems to iterate toward a
glass phase with quasi-long range order characterized by $(J^{*},A^{*})$ where the
fixed point values $J^{*}$, $A^{*}$ are finite and depend on $\alpha$.
This is consistent with recent analytic studies \cite{xyrdm-nsk,xyrdm-kn,xyrdm-s,xyrdm-cf}.
On the other hand, systems with strong disorder, $\alpha > \alpha_c$,
seem to flow to a disordered fixed point $(J^{*},A^{*})=(0,1/4)$ which
corresponds to a non-superconducting glass.
\newline\indent
In $3D$, for $\alpha<\alpha_{c}\approx 0.57$, the system flows to a
strong coupling limit $J^{*}=\infty$. The disorder strength $A(L)$
appears to flow to a finite fixed point value $A^{*}$ which depends on $\alpha$.
However this is not conclusive from our simulations as only very small
values $L=2,3,4,5,6$ are used. This can be interpreted as the zero field
version of a Bragg glass phase. For large disorder,
$\alpha>\alpha_{c}$, $(J(L),A(L))$ seem to iterate to
their maximum values of $(\infty,1/4)$ corresponding to
the gauge glass fixed point. It seems that, in the absence of screening,
$\lambda\rightarrow\infty$, there are two glassy superconducting phases at $T=0$ which, in
an applied magnetic field, correspond to a Bragg glass \cite{giamarchi} for
$\alpha<\alpha_{c}$ and to a vortex glass \cite{fisher:89} for
$\alpha>\alpha_{c}$.

\section{Discussion and Summary}
In this paper, we re-investigate the possibility of an ordered phase
at small but finite temperature $T$ by a numerical domain wall
renormalization group method in a disordered $XY$ model in $2D$ and $3D$
described by the Hamiltonian of Eq. (\ref{eq:h}) in the Coulomb
gas representation. For the $\pm J$ $XY$ spin glass in $3D$, our simulations
yield the spin glass stiffness exponent$\theta_{s}^{BT}\approx+0.10$ which
suggests its lower critical dimension is $d_{l}<3$.
This value of $\theta_{s}^{BT}$ is very different from existing estimates of
the chiral glass stiffness exponent in $3D$ $\theta_{c}\approx +0.47$
\cite{sg-dw-k} and $\theta_{c}\approx +0.56\pm 0.18$ \cite{sg-dw-mg}.
The difference between $\theta_{s}$ and $\theta_{c}$ seems to support
the decoupling of two degrees of freedom in $3D$.
For the gauge glass, we estimate the stiffness exponent
$\theta_{s}^{BT} = -0.36\pm 0.01$ in $2D$ and $\theta_{s}^{BT} = +0.31\pm 0.01$
in $3D$, which are considerably larger than all earlier estimates.
The latter value is consistent with $T_{c}/J \sim {\cal O}(1)$ from
finite temperature MC studies \cite{gg-hs,gg-rtyf,gg-wy}
and also strongly suggests $d_{l}<3$. The results for the $XY$ spin glass
in $3D$ are consistent with spin glass order at $T>0$ which is in contradiction
with all other studies \cite{sg-mc-kt,sg-dw-kt,sg-dw-k} except one \cite{sg-dw-mg}.
\newline\indent
We also studied the effects of varying the disorder strength. In $2D$, our simulations
imply that weak disorder is marginal \cite{xyrdm-nsk,xyrdm-kn,xyrdm-s,xyrdm-cf} and a
system with strong disorder flows to a disordered fixed point. There is no sign
of a re-entrant transition in our simulations. In $3D$, weak disorder
has little effect and the system flows to an ordered phase which is the zero field
analogue of a Bragg glass \cite{giamarchi}. For strong disorder, the system seems
to flow to a gauge glass fixed point. The disagreement between
the stiffness exponent $\theta_{s}^{BT}$ and previous estimates
is because these measure $\theta_{s}^{RT}$ whose meaning is less clear.
The quantity $\Delta E^{RT}(L)$ seems more likely to suffer from large corrections
to scaling as seen in Fig. (\ref{fig:gg3d}), especially for the small system sizes
$L$ which are possible to simulate at present. However, we conjecture that both
measurements would coincide if {\it much} larger values of $L$ could be reached.
Since our simulations are also limited to very small sizes $L$, it is not possible
to draw any definite conclusions from them and more studies are needed to settle
these problems in random systems more satisfactorily.
\newline\indent
One interesting conclusion we can reach concerns the Bragg and
vortex glass states in disordered superconductors in an applied
magnetic field. Recently, one of us \cite{giardina} studied the model
of Eq. (\ref{eq:h}) in the strong screening limit $\lambda\rightarrow 0$
in $3D$ and found that two phases exist at $T=0$ in the presence of an applied
external field. The low field, small disorder phase has a well ordered vortex line
lattice as a ground state with stiffness exponent $\theta_{s} = +1.0$,
whereas the high field, large disorder ground state is a disordered entangled
vortex configuration with $\theta_{s}\approx -1.0$ \cite{kisker,giardina}.
We identify the low field state as a Bragg glass and the high field state as
a disordered entangled vortex liquid. In this limit, the evidence is strongly in
favor of a direct, disorder or field driven transition from a superconducting
Bragg glass to a normal non-superconducting phase. This scenario seems
to be favored by recent experiments.
\newline\indent
In the absence of screening of the
vortex-vortex interactions, the picture which results from this work
is somewhat different, although the studies here are all done in
zero applied field. One may argue that increasing the disorder is
equivalent to increasing the field at fixed disorder. At low field or
low disorder, the ground state is a Bragg glass with $\theta\simeq +1.0$,
exactly as with screening. Without screening, the main difference is
that the high field, large disorder, phase is a true vortex {\em glass}
with stiffness exponent $\theta\simeq +0.30$, as proposed by Fisher
et al. \cite{fisher:89}. We tentatively conclude that a true
superconducting vortex glass phase does not exist in $3D$ {\em except} in the
absence of screening ($\lambda\rightarrow\infty$). Our understanding of
the experimental consequences for real systems with {\em mesoscopic}
penetration depths $\lambda\sim {\cal O}(10^{3})$\AA \, is lacking and may be
of some interest.
\newline\indent
  Computations were performed at the Theoretical Physics
Computing Facility at Brown University. JMK thanks A. Vallat and B.
Grossman for many discussions about spin glasses, best twists etc.

\end{document}